\documentclass{aastex}
\usepackage{epsfig}
\usepackage{graphicx}
\usepackage{latexsym,multirow}
\usepackage{latexsym,multirow}
\usepackage{slashbox}

\newcommand{\bq}{\begin{equation}}
\newcommand{\eq}{\end{equation}}

\shorttitle{Stellar Mass and Star Formation Rate}
\shortauthors{Mobasher {\it B. Mobasher et al}}

\begin{document}

\title{Relation Between Stellar Mass and Star Formation Activity in Galaxies$^1$}
\author{Bahram Mobasher\altaffilmark{2};
Tomas Dahlen  \altaffilmark{3};
Andrew Hopkins \altaffilmark{4};
Nick Z. Scoville  \altaffilmark{5};
Peter Capak  \altaffilmark{5};
R. Michael Rich \altaffilmark{6};
David B. Sanders  \altaffilmark{7};
Eva Schinnerer \altaffilmark{8};
Olivier Ilbert \altaffilmark{7};
Mara Salvato  \altaffilmark{5};
Kartik Sheth\altaffilmark{9}
}

\altaffiltext{1}{Based on observations with the NASA/ESA {\em
Hubble Space Telescope}, obtained at the Space Telescope Science
Institute, which is operated by AURA Inc, under NASA contract NAS
5-26555; also based on data collected at : the Subaru Telescope, which is operat
ed by
the National Astronomical Observatory of Japan; the XMM-Newton, an ESA science m
ission with
instruments and contributions directly funded by ESA Member States and NASA; the European Southern Observatory under Large Program 175.A-0839, Chile; Kitt Peak 
National Observatory, Cerro Tololo Inter-American
Observatory, and the National Optical Astronomy Observatory, which are
operated by the Association of Universities for Research in Astronomy, Inc.
(AURA) under cooperative agreement with the National Science Foundation; 
the National Radio Astronomy Observatory which is a facility of the National Sci
ence 
Foundation operated under cooperative agreement by Associated Universities, Inc 
; 
and and the Canada-France-Hawaii Telescope with MegaPrime/MegaCam operated as a
joint project by the CFHT Corporation, CEA/DAPNIA, the National Research
Council of Canada, the Canadian Astronomy Data Centre, the Centre National
de la Recherche Scientifique de France, TERAPIX and the University of
Hawaii.}  
\altaffiltext{2}{Department of Physics and Astronomy, University of California, Riverside, CA 92521 USA}
\altaffiltext{3}{Space Telescope Science Institute, 3700 San Martin
Drive, Baltimore, MD 21218 USA}
\altaffiltext{4} {School of Physics, University of Sydney, NSW 2006, Australia}
\altaffiltext{5}{California Institute of Technology, MC 105-24, 1200 East
California Boulevard, Pasadena, CA 91125 USA}
\altaffiltext{6}{Department of Physics and Astronomy, University of California, Los Angles, CA 90095 USA}
\altaffiltext{7}{Institute for Astronomy, 2680 Woodlawn Dr., University of 
Hawaii, Honolulu, Hawaii, 96822 USA}
\altaffiltext{8}{Max Planck Institute for Astronomy, Konigstuhl 17, 
69117 Heidelberg, Germany}
\altaffiltext{9}{Spitzer Science Center, California Institute of Technology, Pasadena, CA 91125 USA}

\begin{abstract}

For a mass-selected sample
of 66544 galaxies with photometric redshifts ($z_{phot}$)
from the Cosmic Evolution Survey (COSMOS), we examine
the evolution of star formation activity as a function of stellar mass
in galaxies.
We estimate the cosmic star formation rates (SFR) over the range
$ 0.2 < z_{phot} < 1.2$, using the rest-frame 2800\,\AA \, flux  
(corrected for
extinction). We find the mean SFR to be a strong function of the galactic  
stellar mass
at any given redshift, with massive systems (log$(M/M_\odot) > 10.5$)  
contributing
less (by a factor of $\sim$ 5) to the total star formation rate density
(SFRD).

Combining data from the COSMOS and Gemini Deep Deep Survey (GDDS),
we extend the $SFRD-z$ relation as a function of stellar mass to $z 
\sim 2$. 
For massive galaxies, we find a steep increase in the $SFRD-z$ relation
to $z\sim 2$; for the less massive systems, the SFRD which also  
increases from z $= 0$ to
1, levels off at $z\sim 1$. This implies that the massive systems have had
their major star formation activity at earlier epochs ($z> 2$) than 
the lower mass galaxies.

We study changes in the SFRDs as a function of both redshift and 
stellar mass for galaxies of different spectral types. We find that the
slope of the {\it SFRD-z} relation for different spectral type of galaxies
is a strong function of their stellar mass. For low and intermediate mass
systems, the main contribution to the cosmic SFRD comes from the star-forming
galaxies while, for more massive systems, the evolved galaxies are the
most dominant population.

\end {abstract}


\section {Introduction}

Among the most outstanding issues in observational astronomy today is
understanding the physics of the star formation process and the parameters 
responsible for it. The cosmic evolution of the star formation rate (SFR),
for example, provides clues towards the assembly of mass and the development 
of 
the Hubble sequence, while parameters which govern the star formation activity
are directly responsible for the evolution of galaxies. Despite significant 
progress on this subject in recent years, there are still a number of 
open questions concerning the star formation process in galaxies. 

The main problems in studying star formation
in galaxies can be divided into two general groups: first, 
uncertainties due to dust extinction and variation in properties of different 
star formation diagnostics \citep{Cra:98,Hop:01,Sul:01}; second, lack
of knowledge about the parameters which control star formation activity
in galaxies (Ellison et al 2007; Li et al 2007; Kaufmann et al 2007; Gomez et al 2005). There have been 
major advances in resolving the problems in the first group by comparing
star formation rate (SFR) estimates from different diagnostics 
\citep{Cra:98,Hop:03,Hop:04,dad:07}, investigating the cause 
of their 
difference \citep{Sul:01} and exploring the role of dust extinction 
in SFR measurements \citep{Cal:07,Sul:01}. Due to severe observational
constraints, however, there has been little progress
in addressing problems in the second group. 

\citet{Cow:96} first studied dependence of the specific SFR (SFR per
unit mass) with redshift. They found the star forming galaxies to have
relatively brighter absolute K-band luminosities ($M_K$) at higher
redshifts ($z\sim 1$) compared to those at lower-$z$. As the $M_K$
is analogous to the integrated stellar mass of galaxies, they conclude
that higher mass systems to mainly contribute to the star formation rate 
density (SFRD) at higher redshifts with the main contribution shifting
to lower mass systems at lower-$z$, a phenomenon they called ``downsizing". 
This was confirmed in later studies
of nearby galaxies \citep{Bri:04,Hea:04} and
more distant systems \citep{Jun:05,Zhe:07,Scob:07}.  
Observed changes in ``downsizing" as a function of
redshift reveals that, the most massive galaxies
form first, with galaxy formation proceeding later on to smaller
systems \citep{Jun:05}.

The correlation between stellar mass and SFR in galaxies as a function
of redshift 
provides insights into the integrated SFR over
the history of galaxies and hence, a picture of galaxy evolution over 
cosmic time \citep{Pap:06,Dic:04}. Moreover, study of
these relations for different morphological types of galaxies elucidates
the origin of the Hubble sequence (Abraham et al 2007; Scarlata et al 2007).  

Here, we study  
the relation between SFR with stellar mass and redshift, using a
homogeneous, complete and unbiased sample of galaxies covering a
wide range in all these parameters. We use the extensive multiwavelength 
data from the Cosmic
Evolution Survey- COSMOS (Scoville et al 2007a, Mobasher et al 2007; 
Capak et al 2007a).
The combination of the depth, area and wavelength coverage in COSMOS is
extremely beneficial for such investigations and complements
existing deeper but smaller area surveys 
\citep{Jun:05,Gla:05,Gia:04}. Results from this study are extensively used
in Scoville et al (2007b) to explore the influence of local density enhancement
on star formation activity in galaxies.

Throughout this paper we assume $H_0 =70$ km\,s$^{-1}$\,Mpc$^{-1}$,
$\Omega_\Lambda = 0.70$ and $\Omega_M =0.30$. Magnitudes are given 
in the AB system.

\section {Observations, Photometric Redshifts and Stellar Masses}

We use the photometric data obtained from our imaging  
surveys of the COSMOS field. This includes data in
U (CFHT), BV{\it riz} (Subaru) and K$_s$ (Kitt Peak/CTIO) bands, 
covering the entire 2 sq.\ deg.\ area of the COSMOS field. 
Details of the photometric observations and data reduction are
presented in \citet{Cap:07a} and Taniguchi et al (2007). For the 
present study, we use a subset of the {\it i}-band selected catalog, 
extended over the entire COSMOS field and complete 
to $i < 25$ mag. To this magnitude limit, we derive accurate
photometric redshifts and spectral types (Mobasher et al 2007) 
and can reliably estimate the SFRs. This covers a significantly larger area
with a depth similar to previous studies (see the compilation by
Hopkins and Beacom 2006).

Photometric redshifts for galaxies in COSMOS are calculated using 
the template fitting method incorporating priors, as described in 
Mobasher et al. (2007).   
For each object, we derive the photometric redshift, the redshift probability
distribution and the best-fitting spectral type. The spectral types cover
E, Sbc, Scd and Im templates (\citeauthor{Col:80} \citeyear{Col:80},
extended to the UV and NIR by \citeauthor{Bol:00} \citeyear{Bol:00})
in addition to two starburst templates \citep{Kin:96}. 
We also consider extinction as an independent variable in the SED fits. 

It is possible that by considering extinction as a free parameter
in the fit, we increase the likelihood of degeneracy in the final
result (i.e. redshift, dust extinction and presence of the old population
all make the SEDs redder). This is examined in Mobasher et al (2007) by
studying the accuracy of photometric redshifts by comparing
the photometric ($z_{phot}$) and spectroscopic ($z_{spec}$) redshifts 
for a sample of galaxies in the COSMOS with available such data, 
expressed by the ratio: 
$\Delta_z\equiv\langle|z_{\rm phot}-z_{\rm spec}|/(1+z_{\rm spec})\rangle$. 
The smallest {\it rms} scatter between the estimated photometric
and available spectroscopic redshifts is obtained when  
treating the extinction as a free parameter (see Table 4 and Figure 5 in
Mobasher et al 2007). We find $\Delta_z=0.03 $
with a small fraction ($\sim$2.5\%) of outliers, defined as $\Delta_z>0.3$.  
Since Mobasher et al (2007) use a sub-sample of the COSMOS galaxies with 
available spectroscopy to calibrate photometric redshifts, it is possible that 
the photometric redshift sample is biased against fainter galaxies, for which
spectroscopic data do not exist in COSMOS. To minimize this, 
we constrain the sample to galaxies with $z < 1.10$ as these have 
the most reliable photometric redshifts and least 
number of outliers (Mobasher et al 2007).  

We estimate the stellar mass corresponding to each galaxy, using the 
photometric redshift, the rest-frame V-band absolute magnitudes ($M_V$)
and $M/L_V $ ratio dependent on the spectral type estimated 
from the best fit SED 
for each galaxy, as described in Mobasher et al (2007). 

\section{Star Formation Rates from UV Luminosities}

The SFRs associated with individual galaxies are
estimated using rest-frame 2800\,\AA\ fluxes.
We estimate this for each galaxy, using
photometric data covering UBVr{\em izK} bands. This is done by 
interpolating between the two bands that straddle rest-frame
2800\,\AA\, wavelength, making use of the shape of the best-fit
SED when doing the interpolation. The exception is at low 
redshifts ($0.20 < z < 0.34$), where 
we use the observed U-band magnitude, extrapolated to 2800\,\AA\,  
using the best-fit SED. 
Having the observed apparent flux, K-correction and photometric redshifts
obtained from the best-fit SEDs, we then estimate the absolute flux in 
rest-frame $\lambda=$2800\,\AA. A
detailed description of this technique is presented in \citet{Dah:07}.

\begin{figure}
%\epsscale{0.8}
\includegraphics[angle=0,scale=1]{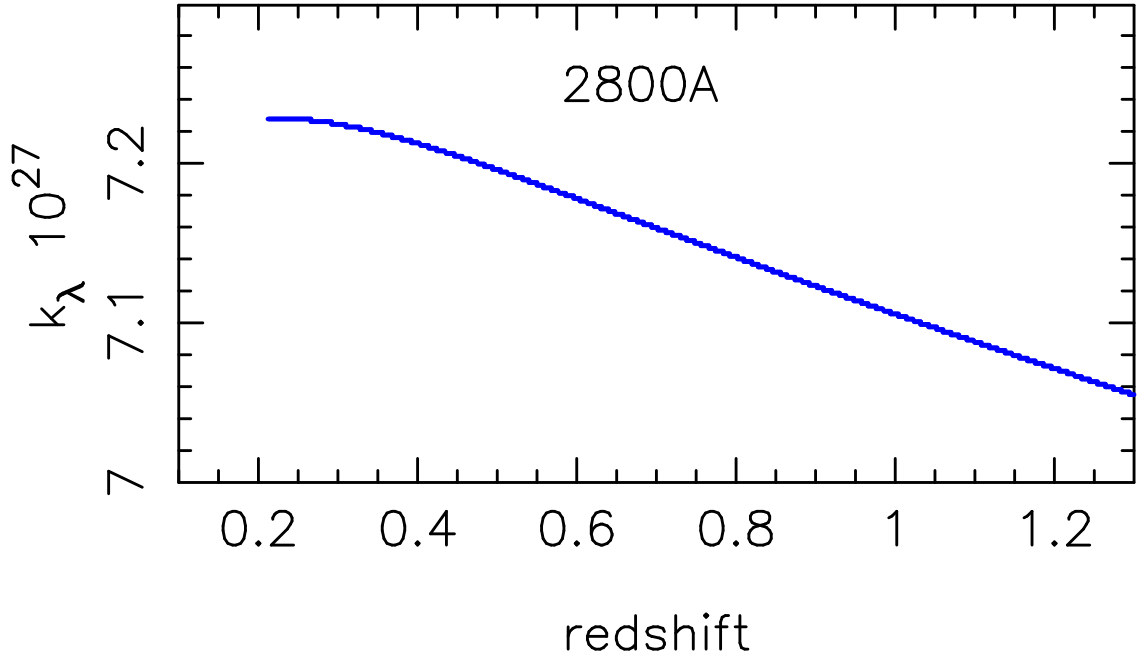}
%\plotone{phot_spec.eps}
\caption{Plot of the conversion factor, $k_{2800}$, between UV luminosity 
(at 2800~\AA) and ongoing SFR as a function of redshift. 
$L_{2800} = k_{2800} (z) \, {SFR \over M_\odot yr^{-1}}$ 
erg s$^{-1}$Hz$^{-1}$.   The
star formation history used is taken from Strolger et al. (2004), while
stellar synthesis models are taken from \citet{BC:03}, assuming
solar metallicity and a Salpeter IMF. The conversion factor is given in 
units of $10^{27}$.
\label{fig1}}

\end{figure}

The UV luminosity at 2800\,\AA\ is mainly produced
by short-lived O and B stars and is therefore closely related to the 
ongoing star formation activity.  
Here, we follow the approach in \citet{Dah:07} to relate the UV flux
to the SFR. This involves using predictions from stellar synthesis models 
and assuming parametric forms for the past star formation history to mimic
the evolution of the SFR with redshift, as summarized below. We
derive the conversion factor between the UV flux
and the ongoing SFR, using the stellar population synthesis code GALEXEV
\citep{BC:03}. We assume solar metallicity and a modified Salpeter IMF
spanning
$0.1 < M/M_\odot < 100$, but having a turnover at low mass \citep{BG:03}.
To model the dependence of SFR on redshift, we use a parametric fit to 
the data from the Great Observatories Origins Deep Survey (GOODS), 
presented in \citet{Gia:04} and parameterized in \citet{Str:04}.
Convolving the past SFR history with the results from the stellar 
synthesis models, we derive a redshift dependent conversion factor,
$k_\lambda (z)$, between the ongoing SFR and UV flux as
\begin{equation}
L_{2800}=k_{2800} (z)\frac{{\rm SFR}}{M_{\odot}~{\rm yr}^{-1}}{\rm ergs}~{\rm 
s}^{-1}
~{\rm Hz}^{-1}
\end{equation}
The dependence of $k_{2800} (z)$~ on redshift is shown in Figure~\ref{fig1}.  
Note that we use the same volume averaged past SFR history to calcluate the 
conversion factor for all galaxies. Therefore, galaxies that have 
SFR histories
significantly different from the average, may be assigned a SFR that differs 
from the actual value. Furthermore, galaxies of different stellar
masses are characterized by different star formation histories (Panter et
al 2007), introducing a mass-dependent bias into the SFR estimates for
individual galaxies. We do not correct the SFR calibration here for this
effect, as this is highly model-dependent and introduces more free parameters
into the analysis. Moreover, any such correction would complicate the
interpretation of a mass-SFR-redshift relation discussed in the following
sections. We show the redshift
dependence of the conversion factor in Figure 1, illustrating only a
weak dependence with a change of only 2\% over the redshift 
interval considered here. This suggests that the dependence on the assumed
SFR history is weak when deriving the SFR. 
While this is an acknowledged source of uncertainty affecting calibration
of the SFR for any individual galaxy, it has no effect on the estimates
of the total SFR density (SFRD). This is because the SFRD is calculated by
integrating over UV luminosity functions estimated from the data, after
scaling the UV luminosity appropriately to SFR using the redshift dependent
conversion shown in Figure 1. We will return to the discussion of the
accuracy of the SFR calibration again in section 5.1, when we compare
our estimated SFRDs with those derived independently and compiled from
literature.

Accurate estimate of dust extinction is essential in correcting the SFRs measured from rest-frame UV (2800 \AA) flux. Using the extinction values estimated for invidual galaxies through the SED fits, and assuming a Calzetti extinction law \citep{Cal:00}, we derive the reddening, $E_{B-V}$, for each galaxy. To examine the accuracy of the $E_{B-V}$ values in this study, we explore their distribution for different spectral types of galaxies. As expected, we find the median extinction ($E_{B-V}$) values to increase in galaxies from old and evolved to young and star-forming spectral types. Furthermore, we estimate the dust corrected SFRDs as presented in section 5.1, with different extinction corrections: the median $E_{B-V}$ values for each spectral type, individual extinction corrections for each galaxy and a constant $E_{B-V}$ value for all the galaxies. We find that the final results do not depend on the specific prescription for extinction correction.

\section{Sample Selection and Completeness}

\begin{figure}
%\epsscale{0.8}
\includegraphics[angle=-90,width=15cm]{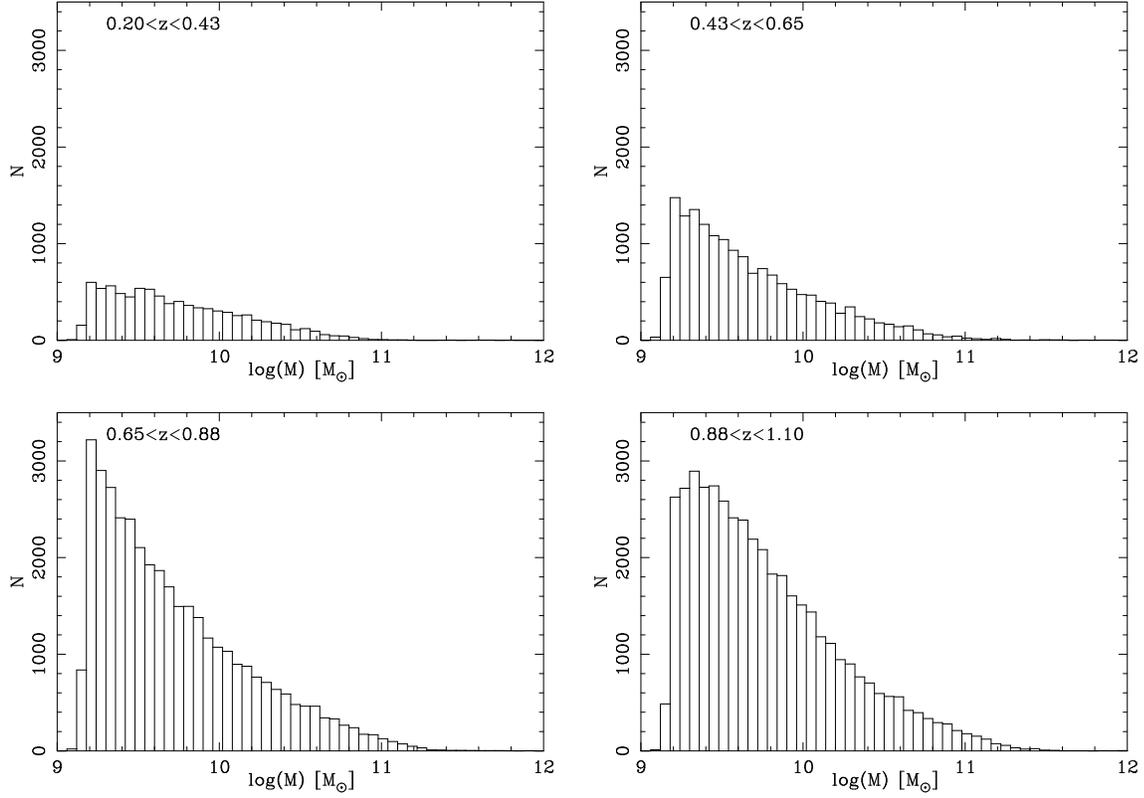}
%\plotone{phot_spec.eps}
\caption{Distribution of the stellar mass of galaxies in different redshift intervals. The redshift bins are adopted following Scoville et al (2007b). 
The mass completeness limits for the COSMOS are estimated and used to select
a survey complete in stellar mass.
\label{fig2}}
\end{figure}

\begin{figure}
%\epsscale{0.8}
\includegraphics[angle=-90,width=15cm]{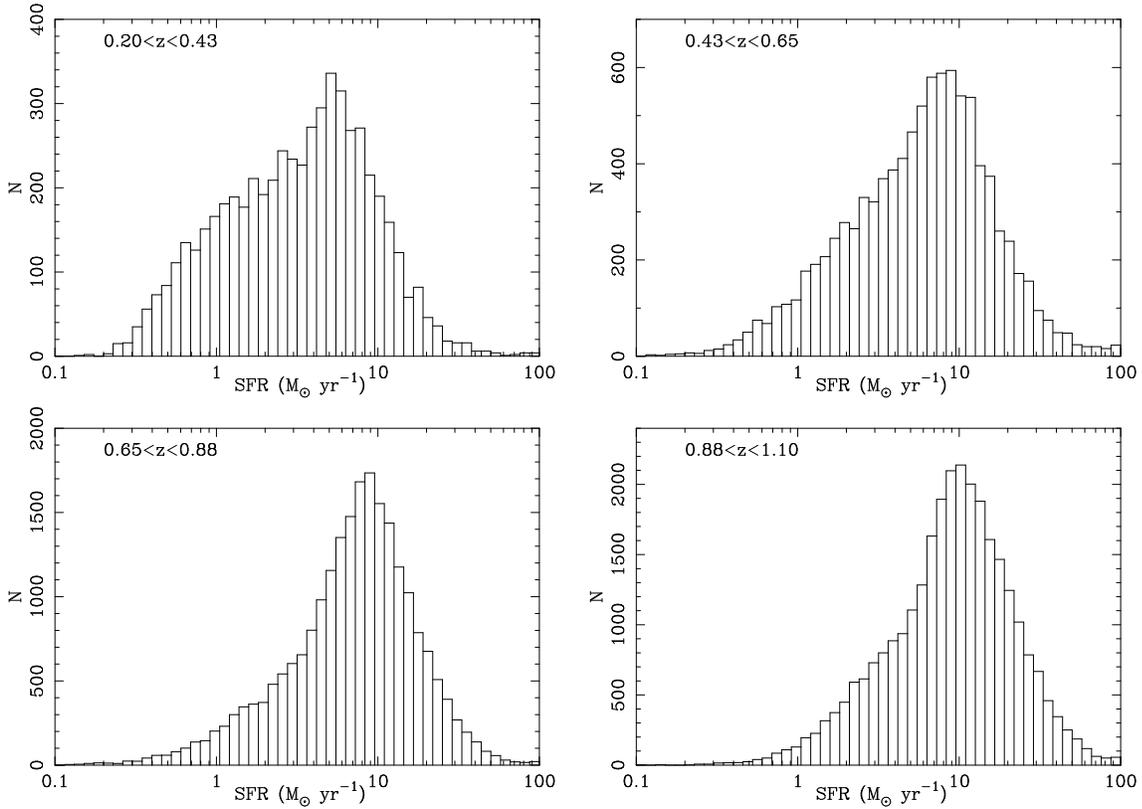}
%\plotone{phot_spec.eps}
\caption{Distribution of the estimated SFRs (corrected for extinction) for
COSMOS galaxies in different redshift intervals. 
\label{fig3}}

\end{figure}

The sample considered in the present study is selected 
in {\em i}-band. This is the filter with the 
deepest COSMOS photometry and provides a
sample most like a mass-selected survey (at least to $z\sim 0.5$). 
Samples of star forming galaxies, selected based on emission lines
([OII], H$\alpha$), broad-band (i.e. UV, U- band), far-infrared or radio 
flux, can be very biased, since they select only a particular class of 
galaxies. This is the main reason for selecting a sample 
resembling a stellar mass-selected survey over much of its 
redshift range. The final sample has: 

\begin {itemize}

\item $i_{AB} < 25$: the majority of the objects fainter
than this limit escape detection at some wavelengths (especially in the 
near-infrared), resulting in poor wavelength coverage of the SEDs and 
thus, uncertain photometric redshifts, SFRs and stellar masses.

\item $0.20 < z < 1.10$. The lower-limit for the photometric redshift 
is adopted to minimize extrapolation to rest-frame 2800~\AA~
for low redshift galaxies, while the upper-limit is selected to avoid
objects with uncertain photometric redshifts (Mobasher et al 2007).

\item $9.5 < \log(M/M_{\odot}) < 11.5$: this allows
a sample complete in terms of stellar mass. (Figure~\ref{fig2} shows the
distribution of stellar mass as a function of redshift). We find that galaxies with redder SEDs (i.e. those with spectral types resembling the early-type systems), are highly incomplete to $\log(M/M_{\odot}) \sim 9.5$ while this is the appropriate completness limit for galaxies with bluer SEDs (corresponding to later-type systems).

\item  0 $M_\odot$/yr $ < SFR_{2800} < 100\ ~ $M$_\odot$/yr: this will exclude
sources with excessive (and probably uncertain) SFRs, which could dominate
the SFR densities, and AGNs wrongly classified as star-forming
galaxies due to their extreme UV flux. 
Figure~\ref{fig3} shows the SFR distribution as a function of redshift. 
Very few galaxies are found to have $ SFR_{2800} > 100 \ M_\odot yr^{-1}$ and therefore, the results are not sensitive to the upper limit for the SFRs.

\item $M_V < -19$: to allow the selection of galaxies from the same part of
the luminosity function at different redshifts and minimize luminosity (stellar
mass) dependent biases. This criterion excludes only an additional 61
galaxies on top of the above criteria, predominantly located in the
middle two redshift bins.

\end{itemize}

After applying the above selection criteria, we have a total 
of 66544 galaxies in our sample. Following Scoville et al (2007b), we divide
the sample into the following redshift bins: $0.20 < z < 0.43$ (5594 galaxies);
$0.43 < z < 0.65$ (9608); $0.65 < z < 0.88$ (22374) and $0.88 < z < 1.10$
(28968). In terms of stellar mass, our combined sample 
is complete to $M\sim 3\times 10^9$ M$_\odot$ over the redshift range
$0.20 < z < 1.1$. 

The fraction of the AGNs in the final sample is $< 1\%$ (a total of 1865 
sources. M. Salvato -private communication), as identified by their X-ray 
flux. We remove all the identified AGNs from the sample. There is a 
possibility that the final sample is still contaminated by a small number
of AGNs. However, while massive galaxies
at $z\sim 2$ could host obscure AGNs (Daddi et al 2007), the fraction of
galaxies at $z < 1$ with strong nuclear activity is small. 
It is also likely that we miss a population of highly  
dusty star-forming galaxies with masses above our stellar mass 
completeness limit. Our estimated global SFRs are therefore likely to be 
lower limits, although our corrections for dust obscuration to the rest-frame UV flux and the technique used to estimate the total SFRDs are
expected to account for the majority of such missing systems.

\section{Results}

\subsection{Measurement of the Star Formation Densities}

The aim of this study is to explore the effect of stellar mass on the
star formation activity in galaxies and to follow its behavior as a function
of look-back time. Previous such studies have either been too shallow
(i.e. 2dF and SDSS- Gomez et al 2005) or, covered only small volumes 
with very few sources
\citep{Jun:05}, therefore, suffering from cosmic variance. 
The present study, based on the COSMOS field, uses data in a wide (2 sq.\ deg) 
area to medium depth, providing a highly homogeneous sample in the
range $ 0.20 < z < 1.10$, with well-known selection bias. 

\begin{figure}
%\epsscale{0.8}
\includegraphics[angle=-90,width=15cm]{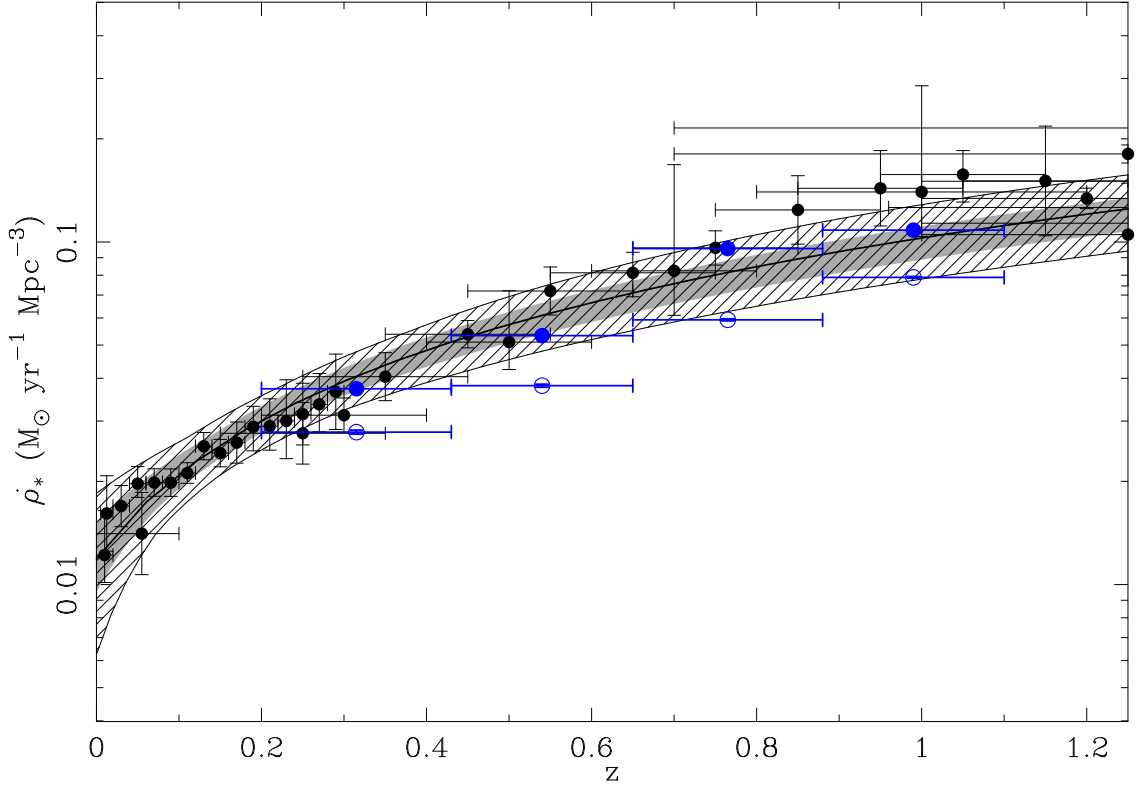}
%\plotone{phot_spec.eps}
\caption{The extinction-corrected SFR densities, $SFRD_{2800}$, {\it vs.} 
redshift, derived from COSMOS galaxies are compared with other 
independent estimates of the SFRD. Total SFRDs are estimated
from integrating the rest-frame UV luminosity function (filled circles)
or are estimated by summing over the SFRs for individual galaxies
(Open circles). Vertical errorbars correspond to
Poisson counting statistics, and are all smaller than the symbol size
for the COSMOS galaxies (blue symbols). The filled black symbols are the
compilation of data from literature (Hopkins \& Beacom 2006), 
reduced to the same cosmology used here.
\label{fig4}}
\end{figure}

We estimate the star formation rate densities (SFRD) in four redshift
intervals: $0.20 < z < 0.43$; $0.43 < z < 0.65$; $0.65 < z < 0.88$ and 
$0.88 < z < 1.10$. In each redshift interval the SFRD
is estimated in two ways: (1) by summing over the extinction corrected SFRs 
for galaxies above the stellar mass completeness limit in each 
redshift interval and normalizing them to the volume corresponding to 
that redshift slice; (2) by integrating the rest-frame UV luminosity function
of galaxies (corrected for extinction) in the above redshift intervals, 
measuring the UV luminosity densities
and converting them to SFRDs. The SFRDs from the two methods are listed
in Table 1, with the evolution of the SFRD
with redshift, derived from the COSMOS galaxies, presented 
in Figure~\ref{fig4}. This is 
compared with the SFRDs estimated independently from other studies, as  
compiled in \citep{HB:06}. All the data in Figure 4 are scaled   
to a modified Salpeter IMF 
described by \citet{BG:03}.  
Errorbars are measured
assuming Poisson statistics. The SFRDs per redshift bin estimated from 
the sum of the contributions from individual galaxies is lower
than those derived from the rest-frame UV LF (Table 1 and  Figure 4). 
This is due to the absence 
of sources fainter than the flux limit of the survey from this method. 
The agreement between the SFRDs estimated here using the extinction corrected
rest-frame UV LF, and those in the literature
over the redshift range $0.20 < z < 1.10$ is very good, 
given that the SFRDs are measured from different diagnostics, different
surveys and are based on different prescriptions for extinction correction. 
The agreement here lends support to our estimated values for
redshift, SFR and extinction. Because of their selection wavelengths, however, 
all these surveys are biased against very dusty, extremely star-forming
galaxies \citep{Afo:03}, implying the estimated SFRDs in Figure~\ref{fig4}
are likely lower limits. We further discuss this and its implications in section 6.

\subsection{Relation between the SFRDs and Stellar Mass}

\begin{figure}
%\epsscale{0.8}
\includegraphics[angle=-90,width=15cm]{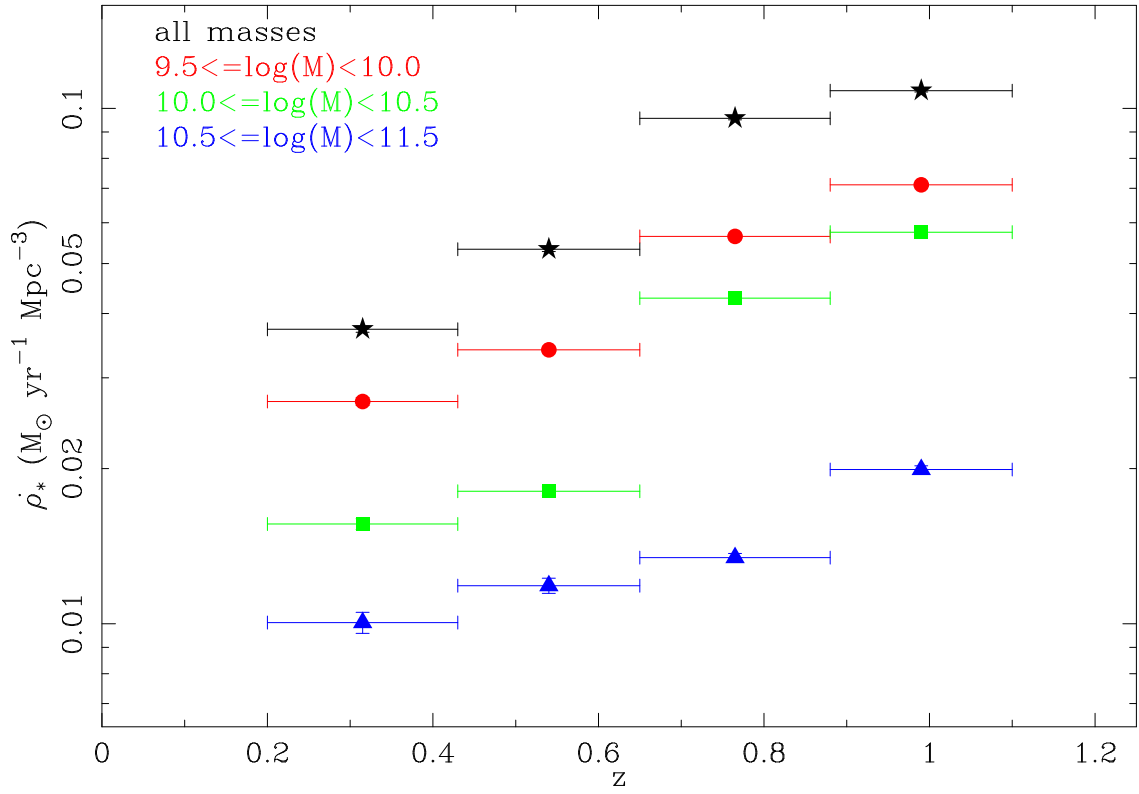}
%\plotone{phot_spec.eps}
\caption{Extinction corrected $SFR_{2800}$ density from COSMOS is plotted 
against redshift in different stellar mass intervals. Errorbars are again
estimated assuming Poisson statistics, and typically smaller than the
symbol size.
\label{fig5}}
\end{figure}

The main result of this paper is presented in Figure~\ref{fig5}, where we
find changes in the SFRDs as a function of redshift and stellar 
mass. We divide the sample into three mass intervals: 
$9.5 < \log(M/M_{\odot}) < 10.0$, $10.0 < \log(M/M_{\odot}) < 10.5$ and
$10.5 < \log(M/M_{\odot}) < 11.5$ in order to have a sufficient number of
galaxies in each stellar mass interval. The points corresponding to 
the lower mass bins are likely to be lower limits due to incompletness. 
Errorbars correspond to Poisson statistics, and are often smaller than
the symbol size. We find that for galaxies in all
mass intervals, there is a clear increase in the SFRDs with redshift, changing
by a factor of 2 to 4 over the redshift range considered here. Moreover, 
at any given redshift, more massive galaxies ($\log(M/M_{\odot}) > 10.5$)
contribute less (by a factor of 5) to the total SFRD. This is because
they have already gone through their intensive star formation phase
and have built up their mass. Results
from Figure~\ref{fig5} also imply that, while massive galaxies 
continue to undergo modest 
star formation activity to $z\sim 1$, they have 
acquired most of their mass before this redshift (i.e. $z>1$).  
The results here are consistent with the ``downsizing"
picture of galaxy formation, where the SFR changes from high-mass 
to low-mass systems (\cite{Cow:96,Jun:05}).

The mass-dependent evolution of the SFRDs show progressively
smaller contributions to the total SFRD (including all masses)
as the SFRD declines to low-$z$. The fraction of the total
SFRD, however, is more or less constant over the redshift range
probed for each mass bin (i.e. at any given mass bin, the ratio of
the SFRD in that bin to the total SFRD does not change over the 
redshift range considered here). This result differs significantly
from that found by Seymour et al (2008) for the evolution
in the high-luminosity end of the luminosity function,
where very strong mass-dependent evolution is seen for
this fractional contribution to the total SFRD. This result
is likely to be due to the different region of the luminosity
function sampled by different selection effects arising
at different wavelengths. Furthermore, the radio selected sample used in
Seymour et al (2008) is mainly 
dominated by galaxies from the high-end of the luminosity function while
the optically selected sample here is more typically dominated 
by $L^*$ galaxies.

Using a sample of galaxies detected at 24 $\mu$m wavelength, Zheng et al (2007)
studied the evolution of the SFRD with redshift as a function of stellar mass. 
Their sample covers the same redshift range as the present study and uses a 
``dust-free" measure of the SFR by combining the UV and total infrared flux
(8-1000 $\mu$m) for individual galaxies. 
Therefore, by comparing our results with those in Zheng et al (2007), 
one could quantify possible dust-induced biases in Figure 5, despite the very
different selection criteria used for these samples. The results from
the two studies are compared in Figure 6, where they are divided into the same
mass intervals. Over the redshift range covered by these studies, the rate
of change of the SFRDs with redshift (i.e. the slope of the SFRD-$z$ relations)
is the same, independent of the stellar mass. For the intermediate mass 
galaxies with
$10^{10.19} < M/M_\odot < 10^{11.19}$, there is excellent agreement
between the the two studies. However, for the lowest mass bin at
$10^{9.5} < M/M_\odot < 10^{10.19}$, the SFRD estimated here is significantly
higher (by a factor of 5) at all redshifts. This discrepancy is not caused by 
dust extinction in our estimate of the SFRD compared to Zheng et al (2007), 
as this would have led to a relatively smaller SFRD 
(with respect to Zheng et al) for our sample. 
This is likely a consequence of both incompleteness and selection
effects in the lowest mass bin of Zheng et al (2007). Their method of
infrared (24\,$\mu$m) stacking, based on the optically selected sample
of COMBO-17 sources in this bin, is biased against optically faint
low-mass galaxies. These are exactly those systems that are likely (by
virtue of dust obscuration) to be, on average, brighter in the infrared
wavelengths than that measured by the stacking result from the optically 
detected sources. 
This causes a bias, leading to an underestimation in the inferred SFRD
for their lowest mass bin. Moreover, the inclusion of low-mass red
galaxies in the stacking analysis (as done by Zheng et al 2007) would
serve to reduce the infered average SFR, adding another bias towards
underestimating the true SFRD. A detailed investigation of the extent of
the bias in the low mass bin of Zheng et al (2007) study would be
facilitated by discriminating between blue (star-forming) and red
(non-star-forming) populations. Given the above discussion, it is
likely that their result for the lowest mass bin is an underestimate.
The disagreement seen in the lowest mass bin in Figure 6 is therefore
not surprising. To summarize, while for the intermediate mass galaxies
the SFRDs at different redshifts are in excellent agreement, the observed
discrepancy for the lowest mass galaxies is likely caused by incompletness
and selection bias in the Zheng et al sample. Correcting for these effects 
brings the Zheng et al's result into better agreement with that in the present
study. For the highest mass bin ($M > 10^{11.15}$ M$_\odot$), there is 
serious incompletness in our sample, which explains the relatively lower 
SFRDs measured for galaxies with higher stellar mass from the present study.

We find the number density evolution in the present sample to  
depend on the stellar mass of galaxies. 
The impact of this number density evolution on the SFRD can be
interpreted by considering the ratio of the SFRD 
($\dot{\rho}_*$) to the number density
of star forming galaxies ($\rho_N$), $\dot{\rho}_* / \rho_N$, defined as
the ``characteristic star formation
rate" (cSFR), as a function of mass and redshift. 
The cSFR remains
flat for our low-mass galaxies, at a level of $\approx
1\,M_{\odot}$\,yr$^{-1}$, while dropping by more than an order of
magnitude for high mass systems from $\approx 2\,M_{\odot}$\,yr$^{-1}$
at $z\approx 1$ to $\approx 0.1\,M_{\odot}$\,yr$^{-1}$ at
$z\approx 0.3$. 
Intermediate masses show an intermediate level
of evolution, with their cSFRs comparable to the high-mass objects
at high-$z$, and to low-mass objects at low-$z$. This behavior exhibits
the characteristics of ``downsizing" in galaxy evolution in the sense that
the cSFR decreases with redshift faster for massive galaxies
than for low-mass systems.

Using spectroscopic observations of galaxies over the range $1 < z < 2$, 
selected from Gemini Deep Deep Survey (GDDS), \citet{Jun:05} studied 
changes in the SFRDs with redshift as a function of stellar mass of
galaxies. This provides
a natural extension of the COSMOS study to $z\sim 2$. To study the behavior of
this relation over the redshift range covered by the combined COSMOS and GDDS, 
we divide our sample into the same mass bins as those in \citet{Jun:05} and
compare the results in Figure~\ref{fig7}. For the most massive galaxies
we find the trend in SFRD with redshift to continue to
$z\sim 2$, however for the less massive systems it flattens around
$z\sim 1.1$. This implies that downsizing has been effective at $z > 1$,
and that the massive galaxies were formed before $z\sim 2$ after going
through a period of intensive star formation activity. Results 
in Figures 5 and 7  
confirm that, at any given redshift out to $z\sim 2$, massive galaxies 
contribute less to the total SFRDs than objects with smaller stellar mass. 
However, although the sample here is selected to be close to 
Juneau et al's galaxies (in terms of the range in their stellar mass), 
differences in the selection criteria between the two samples is likely to 
affect the above result.

\subsection{Evolution of the SFRD as a Function of Stellar Mass and
Galaxy Type}

The coverage of the COSMOS field by the Advanced Camera for Surveys (ACS), 
on-board the Hubble Space Telescope (HST), provides high resolution imaging
of galaxies and measurement of their morphologies. Since the
coverage is in only one HST-ACS band, it is not possible
to measure morphologies for galaxies at the same rest-frame wavelength in
all the 4 redshift bins studied here. We therefore use the spectral types
as a proxy to rest-frame ACS morphologies. We have already shown that our
estimated spectral types closely agree with the rest-frame morphologies
in COSMOS (Capak et al 2007b). Nevertheless, one needs to be cautious that the spectral types and morphologies of galaxies are only loosely related. In the following discussion, these terms are used synonymously. Moreover, relations involving photometric redshifts, spectral types and stellar masses of galaxies should be interpreted carefully as these parameters are not independently derived here. 

In Figure~\ref{fig8}, we present the evolution of the SFRD with redshift as
a function of both stellar mass and spectral type of galaxies. 
The spectral types are divided into those with redder SEDs (early types), 
intermediare color SEDs (spirals) and bluer SEDs (late-types and starbursts)- 
(Mobasher et al 2007). 
As with any type classification, there will be some level of misclassification
or systems that do not fall easily into the defined spectral types. 
For the following discussion this limitation must be kept in mind.
In particular, there is some ambiguity regarding irregular galaxies,
which may in some instances fall either in the ``spiral" or ``starburst" 
classes as all these are represented by SEDs similar to those of star-forming
galaxies. Nevertheless, 
we find a clear trend for all the spectral types in Figure 8. Due to serious incompleteness in galaxies with redder SEDs at $\log(M/M_\odot)\sim 9.5$, the relation for these galaxies is not shown in Figure 8. 
For the low 
and intermediate mass galaxies, 
starbursts dominate the star formation activity at high redshifts with 
a comparable contribution from spirals at lower redshifts. 
Also, for the lowest mass systems ($ 9.5 < log(M/M_\odot) < 10$), 
we find a steep increase with redshift in the SFRD for starbursts, compared
to that for the spirals which remain unchanged with redshift. However, 
for the intermediate mass systems ($ 10 < log(M/M_\odot) < 10.5$), the
slope of the SFRD-$z$ relation increases for both spirals and starbursts, 
with the spirals making a relatively higher contribution to the total
SFRD at lower redshifts.   
For higher mass galaxies, the sample is dominated by early-types, with 
the spirals showing a relatively steeper change in their
SFRD with redshift compared to earlier types. 
This is consistent with the expectation from
the downsizing scenario that the most massive systems are dominated
by early-type old galaxies, which have undergone intense star formation activity 
at higher redshifts ($z\approx 1$). An interesting feature in Figure 7
is that the slope of the SFRD-$z$
relation for different spectral types of galaxies are strong functions
of their stellar mass.
 
Perhaps the most revealing population in this diagram are the starbursts.
These are a dominant form of star formation in galaxies at $z\gtrsim 1$,
but their prevalance at lower redshifts declines progressively with
decreasing redshift. This is reflected in the sharp decrease from 
$z\sim1$ to $z\sim 0.20$ of the
starburst contribution to the SFRD with redshift for low and intermediate
mass systems. The fact that most definitions of a starburst favor
low-mass star forming galaxies can be seen in
the relatively large contribution of this population in the lowest-mass
bin, and progressively less towards higher masses. In the high-mass bin,
there are too few starbursts observed to reliably fit luminosity functions,
and the lower limits shown reflect the sum over the detected population only.
Overall, starbursts account for the majority of the star formation in low- to
mid-mass systems at higher redshifts ($z\gtrsim 0.7$), but at lower redshifts
the dominant population shifts to the spiral types in these mass ranges.
The lower the galaxy mass, the lower the redshift to which the starbursts
remain dominant, consistent with the low-mass irregular starburst galaxies
seen in the local universe (Figure 3 in Kennicutt 1998). This is a 
consequence of the change of slope of SFRD-$z$ relations found in
different stellar mass intervals. 

The spiral types are the dominant contributors at low redshift and low and
intermediate mass, although they make a contribution similar to the 
early types 
in the highest mass bin. The evolution of the SFRD for spirals is flatter than
any other classes, which is due to a transformation
from starburst to spiral spectral types with decreasing redshift. We should 
note that, particularly for early types, the lowest mass systems
are expected to form their stars very late in most hierarchical models. 

\section {Discussion}

A number of studies in recent years have explored changes in SFRD
with redshift as a function of the stellar mass of galaxies 
\citep{Gla:05,Jun:05,Zhe:07}.
Here we investigate this relation using a large and homogeneous sample
of galaxies with well-known selection biases, accurate photometric redshifts, 
SFR estimates and stellar mass measurements. 

We find a strong dependence of the SFRD on stellar mass in the range
$0.20 < z < 1.10$, with higher mass
galaxies ($M > 10^{10.5}$ M$_\odot$) contributing less (by an order of 
magnitude) to the global SFRD than lower mass systems at any given redshift. 
However, the rate of evolution of the SFRD with redshift is the same, 
regardless of the stellar mass of galaxies, in agreement with the
independent study by Zheng et al (2007), in which the effect of
dust obscuration in selection of star forming galaxies and measurement 
of their SFRs is taken into account. 

Using the GDDS with spectroscopic redshifts in the range $1 < z < 2$, 
\citet{Jun:05} studied this relation from $z\sim 1$ to $z\sim 2$. The combined 
COSMOS and GDDS data show that for high mass galaxies
($11.06 < log (M/M_\odot) < 11.76 $), the upward trend in the SFRD continues 
to $z\sim 2$, while for
lower mass systems, after a steep increase with redshift, the SFRD flattens
at $z > 1$. This implies that massive galaxies produced most of their stars
by $z\sim 2$ (when the Universe was $\sim 3.5\,$Gyr old), with the
less massive 
galaxies hosting efficient star formation activity only after $z\sim1$. 
These results confirm that the stellar mass of galaxies regulates the relative
contribution of a galaxy to the global SFRD, consistent with the ``downsizing''
scenario for the formation of galaxies. Moreover, we find that
galaxy ``downsizing'' was already occuring
at $z\sim 2$ and has continued to the present epoch. 

The estimated SFRD in this study is likely to be a lower limit due to
selection at optical bands biasing against dusty star-forming galaxies, 
which could significantly contribute to the global SFRD. 
Using a near-IR selected sample with $K_s < 21.5$,
mimicking a mass selected sample complete to $M \sim 10^{10}$ M$_\odot$,
Caputi et al (2006) estimated the space density of Luminous Infrared
Galaxies (LIRGs) and Ultra-Luminous Infrared Galaxies (ULIRGs) in the range
$0.5 < z < 1$, similar to the redshift range covered in the present study. 
They found that 24\% of galaxies with  
$10^{11} $M$_\odot$ $< M < 2.5\times 10^{11}$ M$_\odot$
are either LIRGs or ULIRGs (excluding AGNs), with a negligible fraction 
of LIRGs and ULIRGs to have masses in excess of 
$2.5 \times 10^{11}$ M$_\odot$. Furthermore, 
using the MIPS observations in
the GOODS-S field, Caputi et al (2005) show a very small fraction of LIRGs or
ULIRGs at $0.4 < z < 1$ to have $M < 5\times 10^{10}$ M$_\odot$. Indeed, 
ULIRGs of any stellar mass are found to be very rare below $z \sim 0.5$ 
\citep{Flo:99,Capu:06,LeF:05}.  
The conclusion from these studies is that, the fraction of massive ($M > 10^{11}$ M$_\odot$) LIRGs and ULIRGs at $z < 1$ is very small, implying that these
objects do not significantly contribute to the total SFRD over this redshift 
range. Since the majority of the LIRGs and ULIRGs at $0.5 < z < 1$
are associated with intermediate mass systems, the absence of these galaxies 
from our sample is likely to lead to an underestimation of the contribution 
from these objects (i.e. intermediate mass systems) to the global SFRD. 
Moreover, \citep{LeF:05} found that at $z > 0.6$, dusty starbursts with $SFR > 10$ M$_\odot$ yr$^{-1}$ dominate the cosmic star formation density, with their contribution increasing from 50\% at $z\sim0$ to 80\% at $z\sim1$ (i.e. only 20\% of the SFRD at $z\sim1$ is due to the UV bright galaxies). 
Any selection bias against these galaxies may possibly compromise
our estimate of the SFRD in the range $0.6 < z < 1.2$. We believe that
a possible bias against such galaxies does not affect our
results here. By estimating the SFRDs through integrating the rest-frame
UV(2800 \AA) luminosity functions (corrected for dust
obscuration) over their entire magnitude range, we incorporate a
contribution from galaxies at the faint-end of the UV luminosity function,
where heavily obscured starbursts would lie.
The excellent agreement between our estimated SFRDs from COSMOS and those
based on the sample compiled by Hopkins and Beacom (2006) in Figure 4
supports the fact that the bias against dusty starburst galaxies does
not significantly affect the results in the present study.
We note that Hopkins and Beacom (2006) estimate the total contribution to
the SFRD from their compilation by combining the SFRD inferred from the
UV and far-infrared luminosities. Nevertheless, the limitations imposed
by selection criteria and any biases so introduced remains a major source
of uncertainty in any such study.

Using a sample of 24 $\mu$m selected galaxies in GOODS-S, Daddi et al (2007)
found a roughly linear relation between the stellar mass and SFR densities. 
For a given mass, the SFRD at $z\sim 2$ was larger by factors of 
$\sim 4$ and $\sim 30$ relative to the star formaing galaxies
at $z\sim 1$ and $z\sim 0$ respectively. This is roughly consistent
with our results in the two highest mass bins in Figure 6. In the range
$1 < z < 2$, the observed 24 $\mu$m wavelength corresponds to rest-frame
$8-12\ \mu$m bands, sampling the PAH features which are most sensitive
to the star forming galaxies. The agreement between or results and 
Daddi et al (2007) further implies that the potential selection bias against 
dusty star forming galaxies in our sample is not significantly affecting
the estimated SFR densities.

By dividing our sample into three spectral types; ellipticals, spirals and
starbursts, we study the evolution (with redshift) of the SFRD-Mass
relation for each type. A striking feature, presented in Figure 7, is
that the slope of the SFRD$-z$ relations, found
for different spectral types of galaxies are strong functions of the
stellar mass.  We find the massive galaxies, which are 
dominated by early-types, to significantly contribute to the total SFRD
at earlier epochs ($z\sim 1$). This is expected as more massive galaxies 
are able to support higher rates of star formation than lower mass systems, 
all else being equal.
For the intermediate mass systems, 
the contribution from early-type galaxies to the global SFRD is minimal, with
no significant change with redshift. The relation for lower mass systems, 
which are dominated by spirals and starbursts, show a progressive build up 
of the stellar mass (i.e. increasing SFRD) with cosmic time. 
The results
here imply that the stellar mass of galaxies plays a crucial role in
determining their star formation activity and spectral types and governs
the contribution of individual galaxies to the total SFRD. 

\section {Summary and Conclusion}

Using multiwavelength data from the COSMOS survey we studied the dependence
of the cosmic star formation density on redshift, stellar mass and
spectral types. We explored the main parameters which govern the star formation
process in galaxies. Our main results are summarized as follows:

\begin {itemize}

\item There is a strong dependence of the SFRD on stellar mass of galaxies, 
with the most massive syatems (log$(M/M_\odot) > 10.5$)
contributing least (by a factor of $\sim$5) to the cosmic SFRD 
at any given redshift. 

\item Combining data from the COSMOS and GDDS, we extend the $SFRD-z-$mass
relation to $z\sim 2$. For high mass galaxies, we find a steep increase
in this relation till $z\sim 2$. This implies that the massive galaxies 
seen today, 
went through intensive star formation activity at $z>1$, to generate their
current stellar mass. For the less massive systems, the $SFRD-z$ 
relation flattens at $z\sim 1$, indicating that these systems are currently
undergoing build up of their mass, with most of their stellar mass
formed at lower redshifts ($z < 1$).

\item We study dependence of the SFRD$-z-$mass relation on the spectral
type of galaxies. We find that the slope of the SFRD$-z$ relation
for different spectral types is a strong function of
their stellar mass. For low and intermediate mass systems, the main
contribution to the cosmic SFR comes from the star-forming
galaxies while, for more massive galaxies, older and redder galaxies 
are the most
dominant population, with their contribution to the global SFRD
increasing with redshift.

\end{itemize}

\begin{table*}

\caption[]{SFRDs estimated from the sum of the SFRDs for individual galaxies and from integrating the rest-frame Luminosity functions at four redshift intervals considered here}
\begin{tabular}{ccc}
& &   \\
z & $log (SFRD) $  M$_\odot$/yr/Mpc$^3$ & $ log (SFRD)$ M$_\odot$/yr/Mpc$^3$ \\
  & LF & simple sum \\
& & \\
$0.315 \pm 0.115$ &  $-1.4286 \pm 0.0058$ &  $-1.5550 \pm 0.0058$\\
$0.540 \pm 0.110$ &  $-1.2733 \pm 0.0045$ &  $-1.4196 \pm 0.0045$\\
$0.765 \pm 0.115$ &  $-1.0192 \pm 0.0029$ &  $-1.2273 \pm 0.0029$\\
$0.990 \pm 0.110$ &  $-0.9656 \pm 0.0026$ &  $-1.1033 \pm 0.0026$\\

\end{tabular} \label{table1}
\end{table*}

\begin{figure}
%\epsscale{0.8}
\includegraphics[angle=-90,width=15cm]{fig6.ps}
%\plotone{phot_spec.eps}
\caption{The SFRDs from the COSMOS (this study)- (filled dots) are compared with those from Zheng et al (2007)- (stars). Different colors indicate different 
mass intervals. 
The two samples are binned in the same stellar mass intervals. The SFRDs from
Zheng et al (2007) are estimated using the combined UV and total infrared
(8-1000 $\mu$m) flux of galaxies. Errorbars correspond to poisson statistics.
\label{fig6}}
\end{figure}

\begin{figure}
%\epsscale{0.8}
\includegraphics[angle=-90,width=15cm]{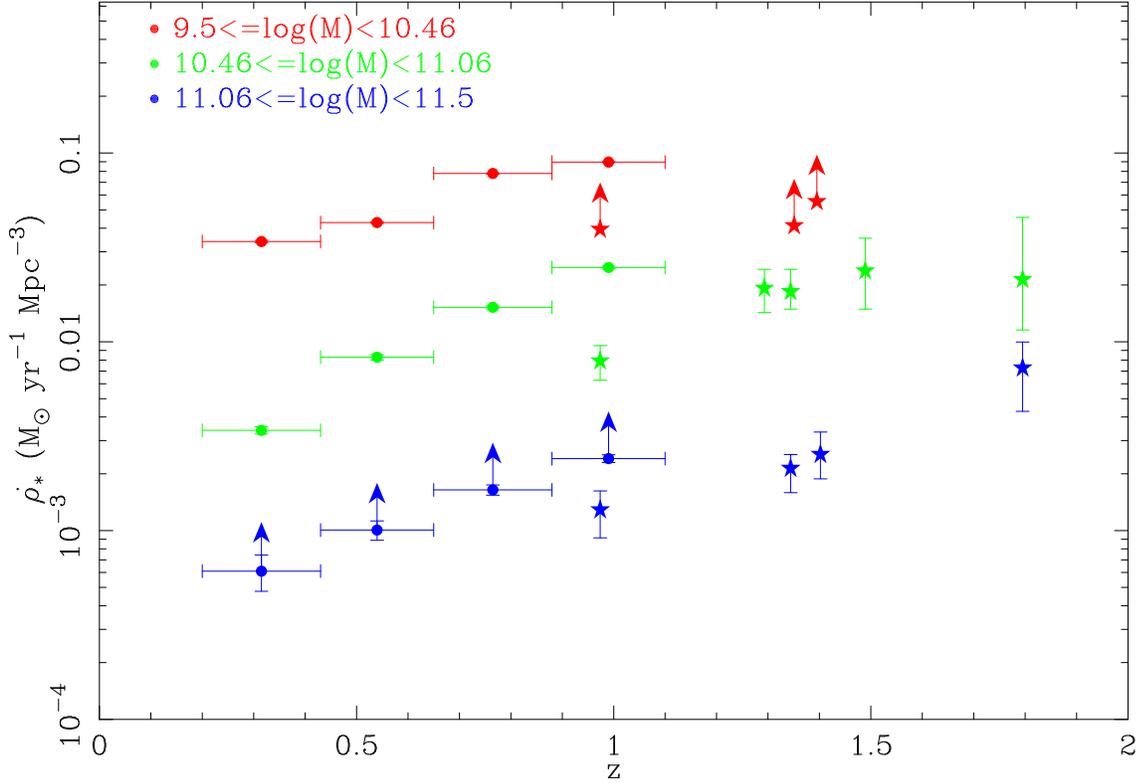}
%\plotone{phot_spec.eps}
\caption{Changes in SFRD with redshift in different stellar mass intervals, 
extended 
to $z\sim 2$ using the \citet{Jun:05} estimates from GDDS (asterix). The SFRDs
from COSMOS are re-binned to the same mass intervals as those in
the \citet{Jun:05} sample. The data from GDDS is reduced to the 
same cosmology and IMF as for the COSMOS.
\label{fig7}}
\end{figure}

\begin{figure}
%\epsscale{0.8}
\includegraphics[angle=-90,width=15cm]{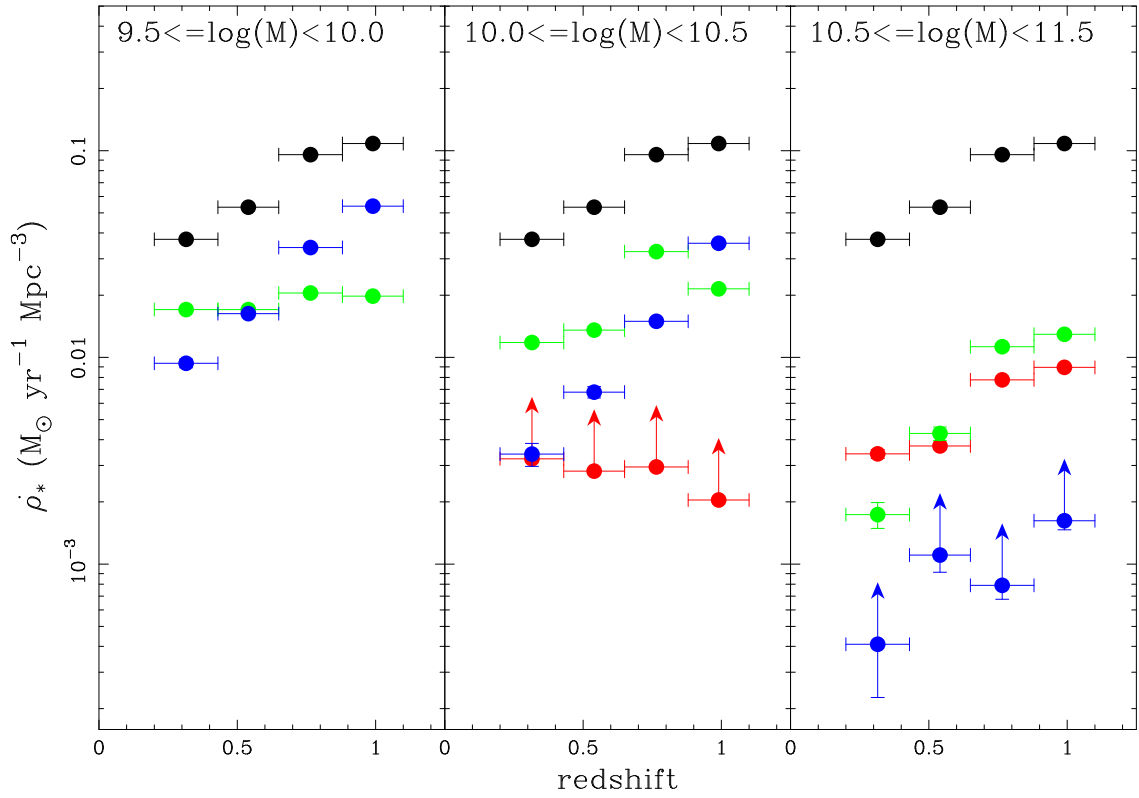}
%\plotone{phot_spec.eps}
\caption{SFRD as a function of stellar mass and galaxy type.
Blue: Starburst; Green: Spiral (Sa-Sd); Red: Early (E/S0); 
Black: all types.
\label{fig8}}
\end{figure}

\end{document}